\documentclass[a4paper,fleqn,usenatbib]{mnras}

\usepackage{newtxtext,newtxmath}

\usepackage[T1]{fontenc}
\usepackage{ae,aecompl}

\usepackage{graphicx}	
\usepackage{amsmath}	

\usepackage{longtable}

\usepackage{pdflscape}
\usepackage{multicol}

\title[Locating the flickering source in polars]{Locating the flickering source in polars}

\author[I{\l}kiewicz et al.]{
Krystian I{\l}kiewicz,$^{1,2}$\thanks{E-mail: kilkiewicz@astrouw.edu.pl}
Simone Scaringi,$^{2}$
Colin Littlefield,$^{3}$
Paul A. Mason$^{4,5}$
\\
$^{1}$Astronomical Observatory, University of Warsaw, Al. Ujazdowskie 4, 00-478 Warszawa, Poland\\
$^{2}$Centre for Extragalactic Astronomy, Department of Physics, University of Durham, South Road, Durham, DH1 3LE, UK\\
$^{3}$Department of Physics, University of Notre Dame, Notre Dame, IN 46556, USA\\
$^{4}$New Mexico State University, MSC 3DA, Las Cruces, NM, 88003, USA\\
$^{5}$Picture Rocks Observatory, 1025 S. Solano Dr. Suite D., Las Cruces, NM 88001, USA
}

\date{Accepted XXX. Received YYY; in original form ZZZ}

\pubyear{2021}

\begin{document}
\label{firstpage}
\pagerange{\pageref{firstpage}--\pageref{lastpage}}
\maketitle

\begin{abstract}
Flickering is a fast variability observed in all accreting systems. It has been shown that in most cataclysmic variables flickering originates in the accretion disc. However, in polars the strong magnetic field of the white dwarf prevents the formation of an accretion disc. Therefore, the origin of flickering in polars is not clear. We analyzed the changes of flickering amplitude with orbital phase in seven polars in order to reveal its site of origin. We show that at least in some polars there are two separate sources of flickering. Moreover, at least one of the sources is located at a large distance from the main source of light in the system.

\end{abstract}

\begin{keywords}
accretion, accretion discs -- methods: data analysis -- (stars:) novae, cataclysmic variables  
\end{keywords}



\section{Introduction}

Flickering is a stochastic variability that can be detected on all timescales -- from milliseconds to hours. It can be present in optical, UV, and X-ray observations and is observed in all objects undergoing accretion, such as young stellar objects, cataclysmic variables, X-ray binaries, and active galactic nuclei \citep[e.g.][]{2015SciA....1E0686S}. While not all forms of stochastic variability are connected to accretion, flickering is recognized by two main features - power spectral density shape is approximately described by a broken power-law \citep[e.g.][]{2004MNRAS.348..783M} and the existence of a linear relationship between the amplitude of flickering and the average flux \citep{2001MNRAS.323L..26U} that is often named the RMS-flux relationship.

Many possibilities have been suggested as to the origin of flickering. The proposed theoretical models for cataclysmic variables included accretion through the hot spot \citep{1971MNRAS.152..219W}, variable transfer trough the boundary layer \citep{1993A&A...275..219B}, flares in the accretion disc atmospheres \citep{1997ApJ...486..388Y}, magnetohydrodynamic turbulence in the accretion disc \citep{1998RvMP...70....1B}, unstable angular momentum transport trough the inner accretion disc \citep{2010MNRAS.402.2567D}, and a fluctuating accretion disc \citep{2014MNRAS.438.1233S}. However, \citet{2005MNRAS.359..345U} showed that flickering is a multiplicative process and cannot be additive, which reduced the number of viable models of flickering. Currently, the most widely accepted model of flickering is a variable mass accretion rate that propagates through the accretion disc \citep[e.g.][]{2019MNRAS.485..260A}.

Irrespective of the physical origin of flickering, the flickering source can be mapped using eclipses \citep[e.g.][]{1996A&A...312...97B,2008ApJ...676.1240B}. All such studies in cataclysmic variables showed that the flickering originates in the vicinity of the white dwarf \citep[e.g.][]{1985MNRAS.216..933H,1996ASSL..208...33B,2003ApJ...584.1027S}. However, flickering can originate in different regions of the white dwarf vicinity in different systems, ranging from the bright spot \citep{1971MNRAS.152..219W},  throughout the whole disc \citep{1983MNRAS.203..909W,1985MNRAS.216..933H} or the inner disc region itself \citep{2010MNRAS.402.2567D}. In fact, two separate sources of flickering in one system could also be present \citep[e.g.][]{2004AJ....128..411B}. 

Polars are cataclysmic variables in which the magnetic field is strong enough to prevent the formation of an accretion disc and synchronize the white dwarf rotation with the orbital period. Accretion onto polars takes place through an accretion stream that is formed along the magnetic field lines. While all the studies of non-magnetic cataclysmic variables showed the accretion disc as a source of flickering, flickering in polars has to originate elsewhere. If the physical location of the flickering source in polars was found, this could serve as a test of the physical processes behind flickering. This is in particular interesting, as it was suggested that flickering should have the same physical origin in every accreting object \citep{2015SciA....1E0686S}. Moreover, flickering in polars seems to have similar statistical properties as in non-magnetic cataclysmic variables \citep[e.g.][]{2010A&A...519A..69A,2021MNRAS.503..953B}. However, in rare instances where flickering was studied in polars, detailed investigation of the dependence of flickering on orbital phase was rarely performed \citep[e.g.][]{2002A&A...394..171H,2018RAA....18...75W,2021MNRAS.503..953B}. Results presented in literature suggest that flickering in polars originate either as a result of accretion of large diamagnetic blobs \citep{2002A&A...394..171H} or from the post shock region close to the white dwarf surface \citep{2021MNRAS.503..953B}.


Here we investigate the dependence of flickering on orbital phase, in order to explore the production site of flickering in polars. In Section~\ref{sec:obs} we present our observations and methodology. The analysis of the dependence of flickering on the orbital phase is shown in Section~\ref{sec:results}.  Finally, we give a summary of our results in Section~\ref{sec:conclusion}.

\section{Observations and methods}\label{sec:obs}

We selected nine polars that were observed by Transiting Exoplanet Survey Satellite (\textit{TESS}; \citealt{2015JATIS...1a4003R}). The polars were chosen by us from a sample of cataclysmic variables that were monitored by us with  \textit{TESS} \citep[e.g.][]{2021MNRAS.503.4050I,2021AJ....162...49L,2021NatAs.tmp..201S}. The objects were observed with 120s cadence. \textit{TESS} observes sections of the sky in sectors corresponding roughly to 28~days. Each sector was inspected visually and each sector in which the star displayed a large long-term variability, such as e.g. changes in mass transfer rate, were removed from the analysis. For this reason, we removed sectors 2 and 27 in case of CW~Hyi as well as sectors 40 and 41 in the case of AM~Her. Moreover, we excluded data of PBC~J0658.0-1746 taken before MJD~59214, where the mass transfer rate in the system seemed to vary. The list of analyzed sectors, starting dates of each sector MJD$_0$ and list analyzed objects is presented in Tab.~\ref{tab:obs}. Each sector was approximately 27 days long. The dates corresponding to each sector can be found on the \textit{TESS} website\footnote{https://tess.mit.edu/observations/}. We employed \textit{TESS} data processed with the SPOC pipeline \citep{2016SPIE.9913E..3EJ}. In order to remove any secular low-amplitude variability from the data we fitted and subtracted a low-order polynomial from data in each sector.

We measured the orbital periods with a Lomb-Scargle periodogram \citep{1976Ap&SS..39..447L,1982ApJ...263..835S}  using a routine from Astropy \citep{astropy:2013, astropy:2018}. The obtained frequencies were than adjusted and their errors were calculated as standard errors of the fitted orbital variability \citep[see e.g.][]{2021MNRAS.503.4050I}. If more than one sector was available, we measured the orbital period using all sectors simultaneously.  However, we used orbital frequencies from literature if a more accurate measurement was available. The orbital phase was chosen arbitrarily so that the zero phase corresponds to the maximum flux.  The full list of employed orbital frequencies is presented in Tab.~\ref{tab:obs}.

\begin{table}
	\centering
	\caption{A list of analysed polars with the \textit{TESS} sectors in which they were observed, the starting date of each sector,  as well as the adapted orbital frequencies.}
	\label{tab:obs}
	\begin{tabular}{ccrr} 
		\hline

 Star & Sectors	& MJD$_0$ & Frequency [c/d]  \\\hline
 BL~Hyi & 1,2,29 & 58325,58354,59088 & 12.6731924(2)  \\
 PBC~J0658.0-1746  & 33$^{\rm *}$  & 59195 & 10.087358$^{\rm 1}$  \\
 CW~Hyi  & 1,28,29  & 58325,59061,59088 & 7.9326874(2)  \\
 MR~Ser & 24,25 &58955,58983 & 12.68890(3)  \\
 AM~Her & 14,25,26  & 58683,58985,59010 & 7.756321$^{\rm 2}$ \\
 CTCV~J1928-5001 & 27 & 59036 & 14.252666(2)$^{\rm 3}$  \\
 VV~Pup &  34 & 59229 & 14.3374(2) \\\hline
	\end{tabular}
\raggedright
\textbf{Notes}: $^{\rm *}$Only data after MJD~59214\\
\textbf{References}: $^{\rm 1}$\citet{2019MNRAS.489.1044B}; $^{\rm 2}$\citet{2020A&A...642A.134S}; $^{\rm 3}$\citet{2005MNRAS.364..565P}
\end{table}

\begin{figure}
\resizebox{\hsize}{!}{\includegraphics{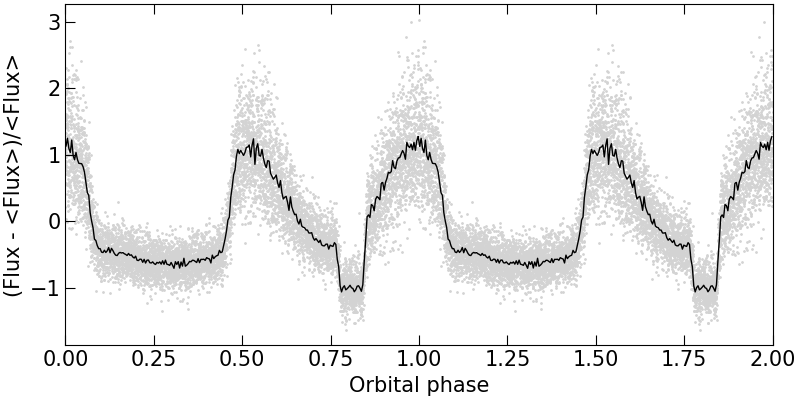}}
    \caption{A phase plot of \textit{TESS} observations of PBC~J0658.0-1746 with the orbital period after subtracting the long term variability (gray points) and the mean orbital variability (black line). }
    \label{fig:example_phased}
\end{figure}

\begin{figure}
\resizebox{\hsize}{!}{\includegraphics{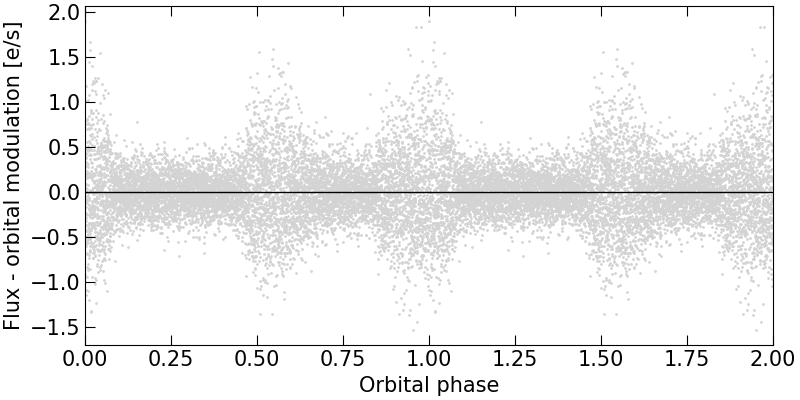}}
    \caption{A phase plot of \textit{TESS} observations of PBC~J0658.0-1746 after removing the orbital variability (gray points). The black line corresponds to zero flux.}
    \label{fig:example_phased_flat}
\end{figure}

\begin{figure}
\resizebox{\hsize}{!}{\includegraphics{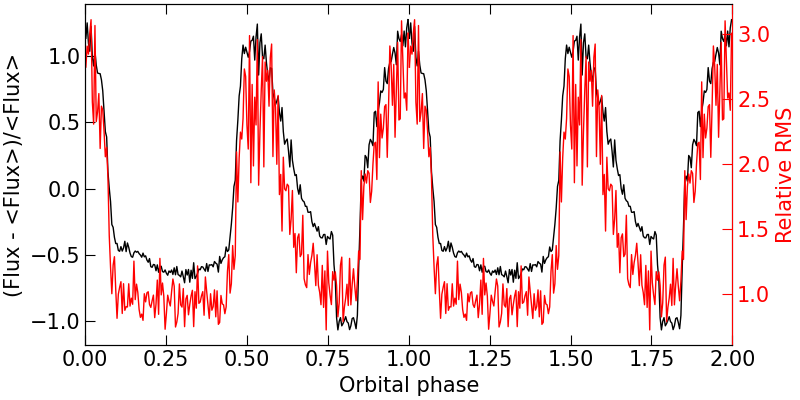}}
    \caption{The mean orbital variability of PBC~J0658.0-1746 (black line) and the RMS of  \textit{TESS} observations scaled by the expected instrumental noise (red line). }
    \label{fig:phased_rms}
\end{figure}

In order to study the evolution of flickering one can use \textit{single} \citep{1996A&A...312...97B,2000A&A...359..998B} or \textit{ensamble} methods  \citep{1985MNRAS.216..933H,1996ASSL..208...33B}. In this study, we used a modified \textit{ensamble} method. Namely, for each sector, we first phased the data with the orbital period and binned the data into 250 bins, equally distributed with the orbital period. Afterwards, we calculated the average flux in each bin, which resulted in an average orbital variability (Fig.~\ref{fig:example_phased}). We then subtracted the mean orbital variability from the data. After this process, the data points represent the deviation from the mean orbital variability that can be plotted against the orbital phase (Fig.~\ref{fig:example_phased_flat}). We calculated the root mean square (RMS) of data points in each of the 250 bins used previously in order to directly compare the dependence of the activity of the star to the changes in its brightness (Fig.~\ref{fig:phased_rms}). The RMS was normalized to the expected level of the instrumental noise provided by \textit{TESS}. Hance, any relative RMS values higher than one cannot be explained by the instrumental noise.

One of the factors increasing the RMS in observations of polars could be quasi-periodic oscillations (QPOs). However, QPOs in polars are observed on a timescale of  0.1-1~Hz \citep[e.g.][and references therein]{2018MNRAS.474.1629B}, which is below our time-resolution. Therefore, the observed changes in RMS should directly correlate with the variability of the flickering amplitude. 


\section{Results}\label{sec:results}

\begin{figure}
\resizebox{\hsize}{!}{\includegraphics{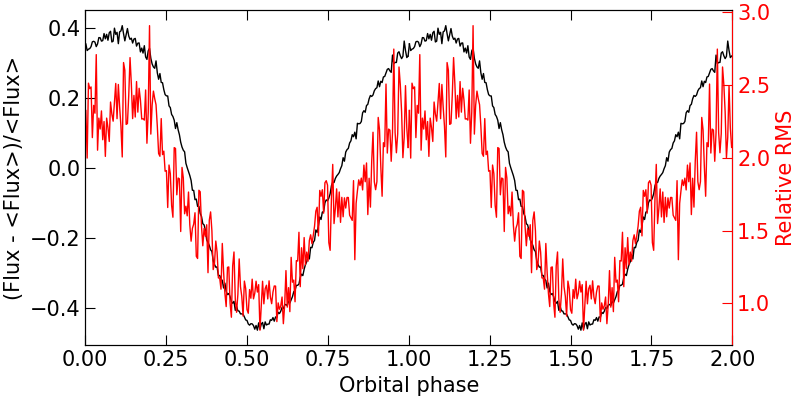}}
\resizebox{\hsize}{!}{\includegraphics{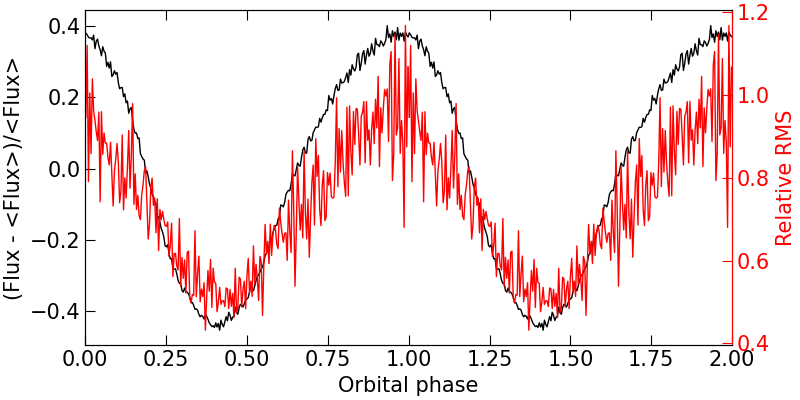}}
\resizebox{\hsize}{!}{\includegraphics{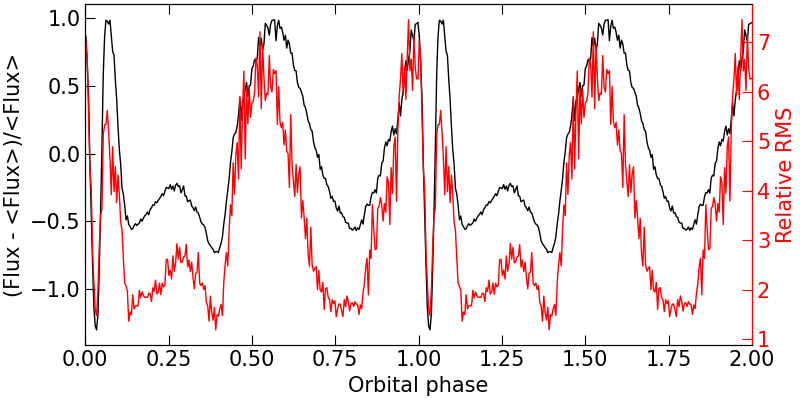}}
    \caption{A mean orbital variability (black line) and a RMS of the \textit{TESS} observations with respect to the mean orbital variability (red line) for the polars with RMS and flux correlated. Top panel: MR~Ser with \textit{TESS} data from sector 24. Middle panel: MR~Ser with \textit{TESS} data from sector 25. Bottom panel: CTCV~J1928-5001 \textit{TESS} data from sector 27. }
    \label{fig:phased_rms_expected}
\end{figure}

While the location of the source of flickering in polars is not known, it has been suggested that it may originate from a cyclotron or reprocessed X-ray radiation at the bottom of the accretion column \citep{2021MNRAS.503..953B}. For this reason, \citet{2021MNRAS.503..953B} suggested that flickering in polars should be correlated with the changes in brightness. Moreover, the same thing should be expected if flickering in polars follows the RMS-flux relation that is universal to flickering in other systems \citep{2015SciA....1E0686S}. In fact, this is what \citet{2021MNRAS.503..953B} observed for V834~Cen. We observed a similar correlation for MR~Ser and CTCV~J1928-5001 (Fig.~\ref{fig:phased_rms_expected}). Notably, CTCV~J1928-5001 is an eclipsing system \citep{2004MNRAS.354..321T,2005MNRAS.364..565P} and the flickering source is eclipsed simultaneously with the main light source from the system (Fig.~\ref{fig:phased_rms_expected}). This is consistent with the model of \citet{2021MNRAS.503..953B}.

\begin{figure}
\resizebox{\hsize}{!}{\includegraphics{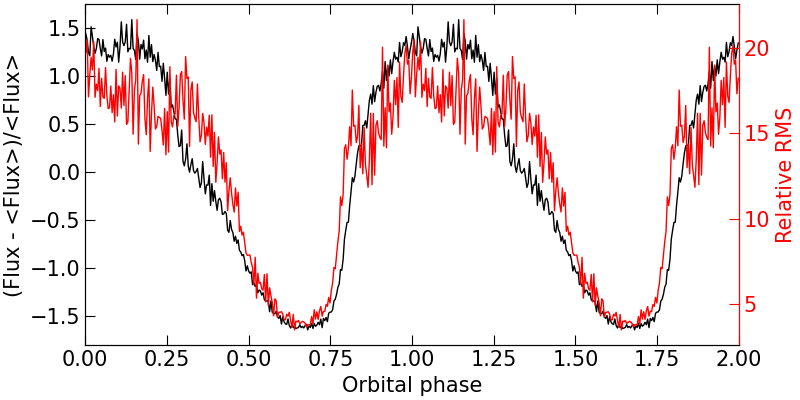}}
\resizebox{\hsize}{!}{\includegraphics{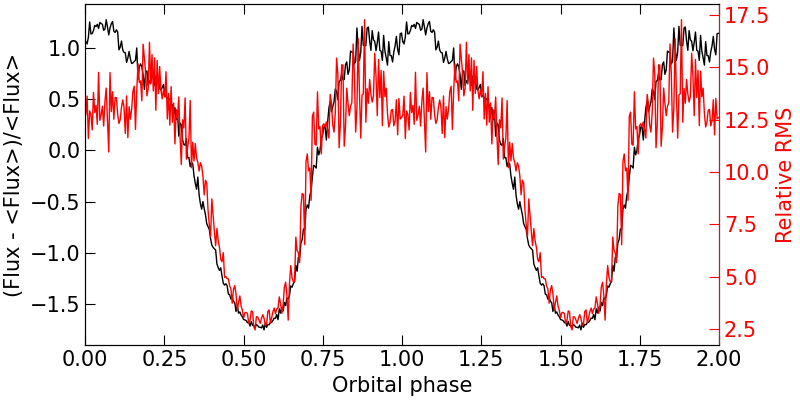}}
\resizebox{\hsize}{!}{\includegraphics{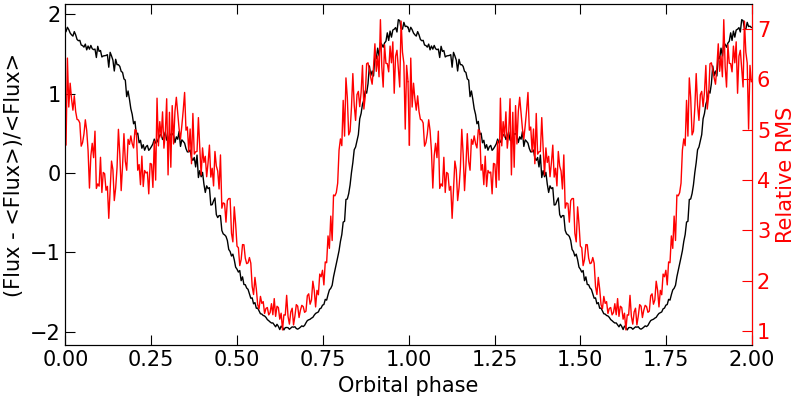}}
    \caption{Same as Fig.~\ref{fig:phased_rms_expected}, but for BL~Hyi and data from \textit{TESS} sector 1 (top panel) sector 2 (middle panel) and sector 29 (bottom panel). }
    \label{fig:phased_rms_blhyi}
\end{figure}

Similar behavior of flickering to MR~Ser and CTCV~J1928-5001 was observed for BL~Hyi during the \textit{TESS} sectors 1 and 2 (Fig.~\ref{fig:phased_rms_blhyi}). However, during \textit{TESS} sector 29 an additional eclipse is visible at orbital phase of $\sim$0.40  (Fig.~\ref{fig:phased_rms_blhyi}). This eclipse was present both in the mean light coming from the system as well as the flickering component. A shallow and contemporary nature of this eclipse suggests that it is due to a partial obscuration of the emitting spot by the accretion stream \citep[e.g.][]{2004MNRAS.354..773R,2014MNRAS.445.1403B,2019MNRAS.489.1044B}. The appearance of a stream eclipse may suggest a significant change in accretion geometry in BL~Hyi between the first two \textit{TESS} sectors and sector 29. Moreover, at orbital phase $\sim$0.28, a new eclipse was present in the flickering component. This eclipse of the flickering component was significantly more pronounced and was not accompanied by a variation of the mean light.

\begin{figure}
\resizebox{\hsize}{!}{\includegraphics{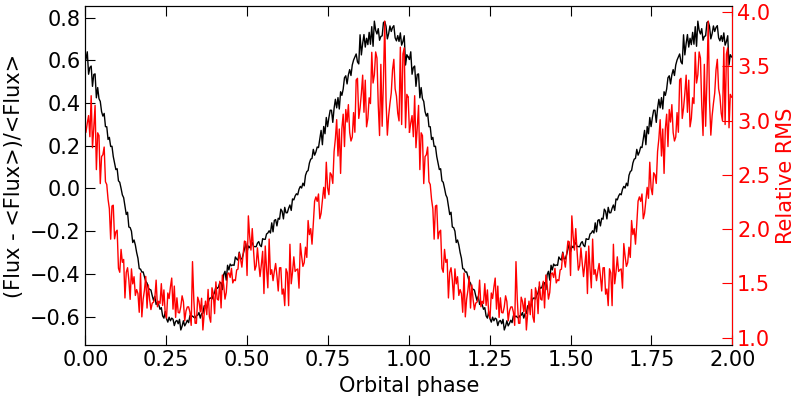}}
\resizebox{\hsize}{!}{\includegraphics{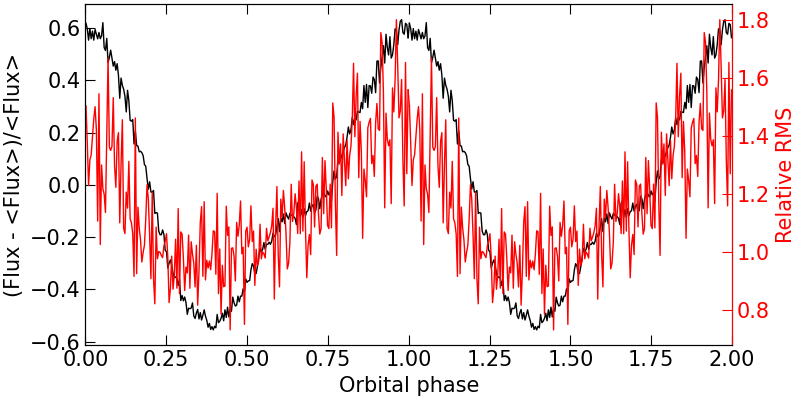}}
\resizebox{\hsize}{!}{\includegraphics{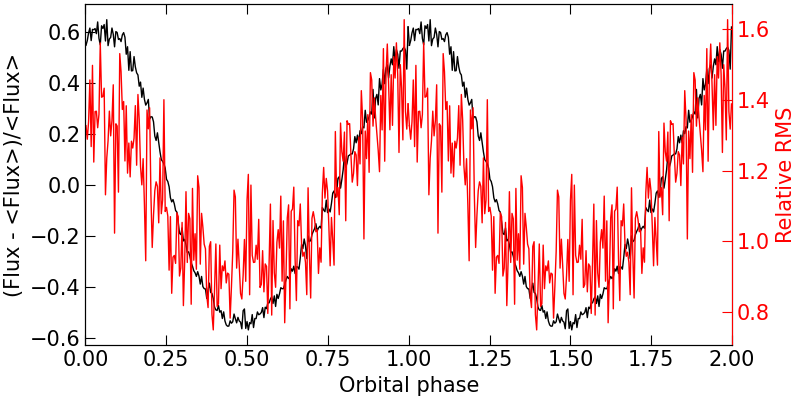}}
    \caption{Same as Fig.~\ref{fig:phased_rms_expected}, but for CW~Hyi and data from \textit{TESS} sector 1 (top panel) sector 28 (middle panel) and sector 29 (bottom panel). }
    \label{fig:phased_rms_cwhyi}
\end{figure}

During  \textit{TESS} sector 1 CW~Hyi appeared to follow the expected correlation between the mean light and flickering strength (Fig.~\ref{fig:phased_rms_cwhyi}). The only exception was an eclipse of the flickering source at orbital phase $\sim$0.0, which was not accompanied by any decrease in the mean brightness. This eclipse is similar to the first eclipse of the flickering source of BL~Hyi during \textit{TESS} sector 29. The eclipse of the flickering source was not present in CW~Hyi during sectors 28 and 29. We note that the flickering was significantly stronger during sector 1 while during sectors 28 and 29 the relative RMS was close to unity, meaning that the eclipse could be hidden in the instrumental noise. The decrease in the flickering amplitude in CW~Hyi during the last two sectors was associated by a decrease in the orbital variability amplitude, which could be due to a decrease in the mass transfer rate.

PBC~J0658.0-1746 is an eclipsing polar \citep{2018AJ....155..247H}. While the eclipse is present in the \textit{TESS} data it is not present in the changes of the flickering amplitude (Fig.~\ref{fig:phased_rms}). Therefore, the behavior of PBC~J0658.0-1746 seems inverse to CW~Hyi and BL~Hyi, where the eclipse of the flickering component was not reflected in changes in the mean brightness.  We note that the RMS level during the PBC~J0658.0-1746 eclipse should be accepted with caution and the RMS likely does not reflect the amplitude of flickering. The high RMS during the eclipse is most likely an artifact due flickering amplitude smaller than the expected accuracy of measurements (Fig.~\ref{fig:phased_rms}). 

\begin{figure}
\resizebox{\hsize}{!}{\includegraphics{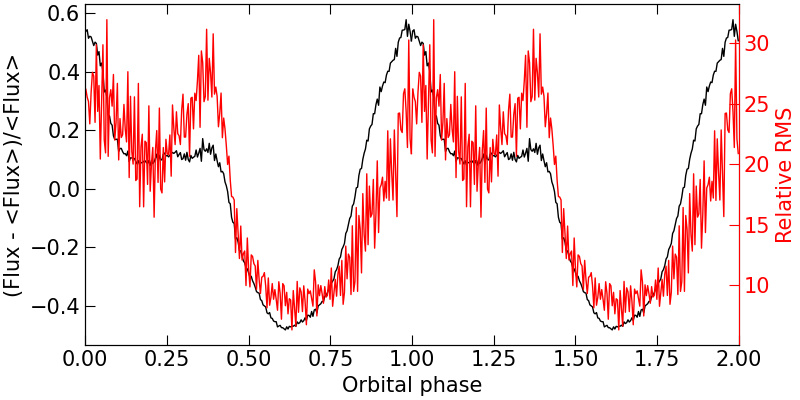}}
\resizebox{\hsize}{!}{\includegraphics{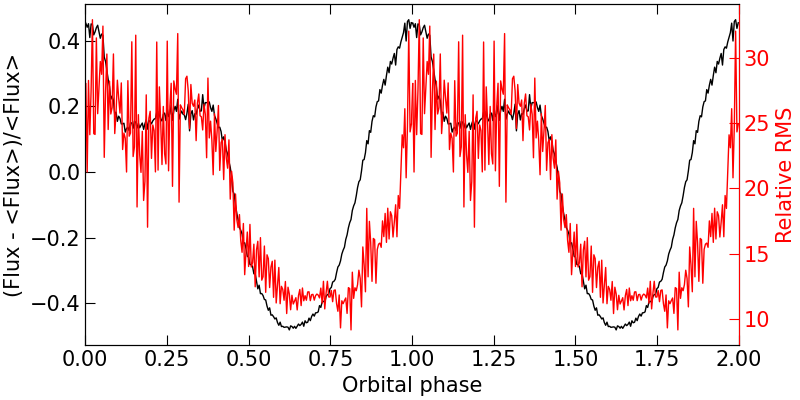}}
\resizebox{\hsize}{!}{\includegraphics{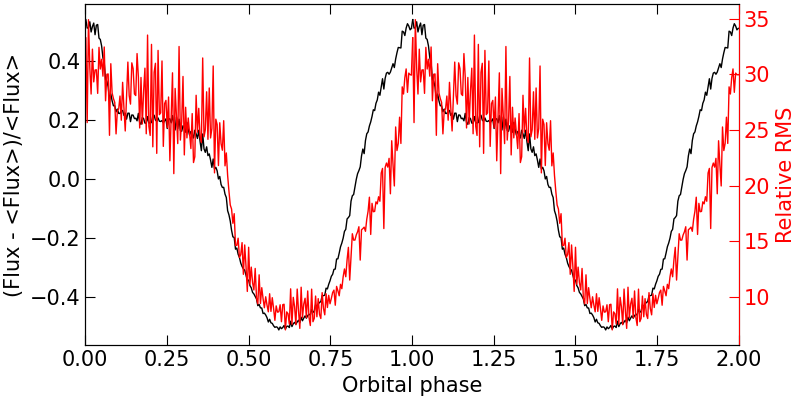}}
    \caption{Same as Fig.~\ref{fig:phased_rms_expected}, but for AM~Her and data from \textit{TESS} sector 14 (top panel) sector 25 (middle panel) and sector 26 (bottom panel). }
    \label{fig:phased_rms_amher}
\end{figure}

AM~Her is the best-studied polar in our sample \citep[e.g.][]{2013ApJ...774..153D}. The mean orbital variability is observed as a peak in the mean light followed by a plateau in all of the observed \textit{TESS} sectors (Fig.~\ref{fig:phased_rms_amher}). Interestingly, the increase in the mean light precedes the increase in the flickering amplitude during all the \textit{TESS} sectors. Conversely to the rise to the peak, during the decrease from the plateau both the mean light and flickering amplitude seem to be correlated. Moreover, it appears that at the end of the plateau in the orbital variability during \textit{TESS} sector 14 a second peak in the flickering amplitude is present. 

\begin{figure}
\resizebox{\hsize}{!}{\includegraphics{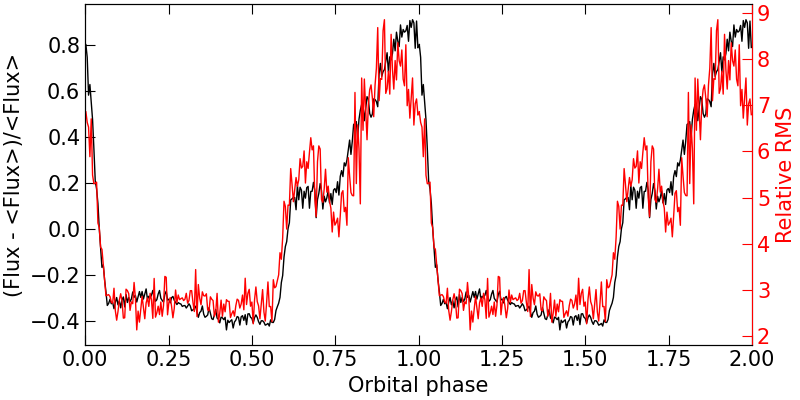}}
    \caption{Same as Fig.~\ref{fig:phased_rms_expected}, but for VV~Pup and data from \textit{TESS} sector 34. }
    \label{fig:phased_rms_vvpup}
\end{figure}

A similar variability to AM~Her is present in VV~Pup. The difference is that in VV~Pup the plateau precedes the peak during the orbital variability (Fig.~\ref{fig:phased_rms_vvpup}).  Similarly to AM~Her during \textit{TESS} sector 14, in VV~Pup a second peak in the flickering component is observed during the plateau. However, conversely to AM~Her both the rise and decrease of the mean light and flickering amplitude in VV~Pup occur simultaneously.

\subsection{The phased RMS-flux relationship}

\begin{figure*}
\resizebox{0.3\hsize}{!}{\includegraphics{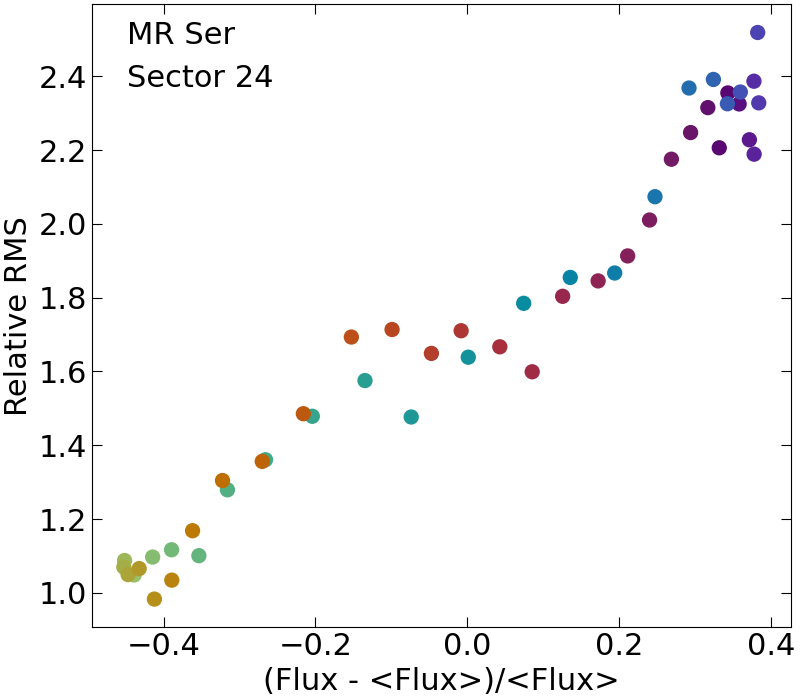}}
\resizebox{0.3\hsize}{!}{\includegraphics{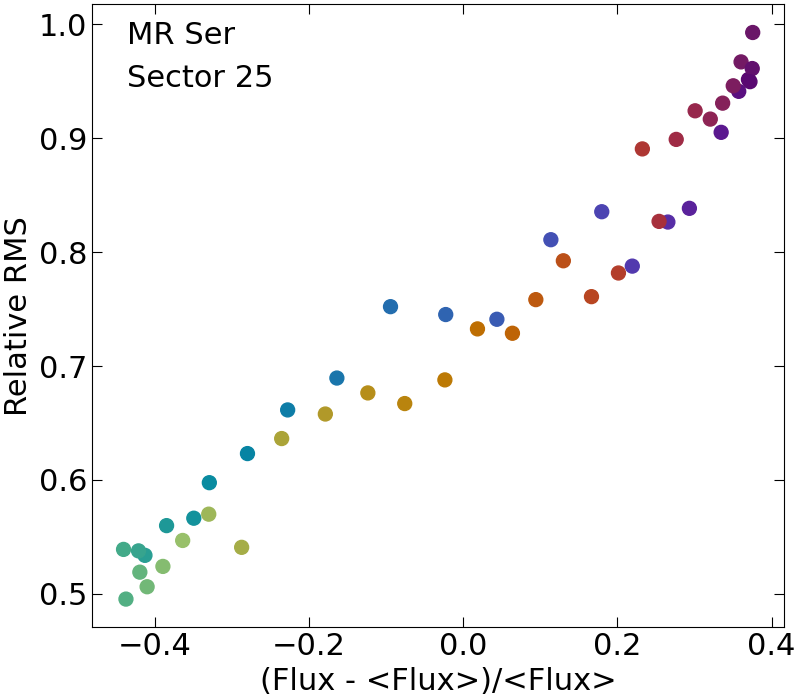}}
\resizebox{0.3\hsize}{!}{\includegraphics{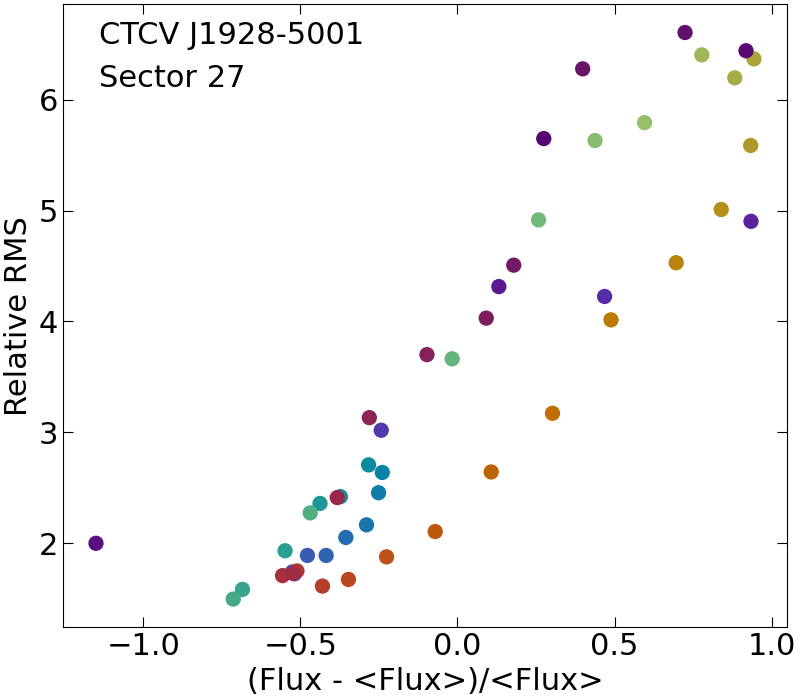}}\\
\resizebox{0.3\hsize}{!}{\includegraphics{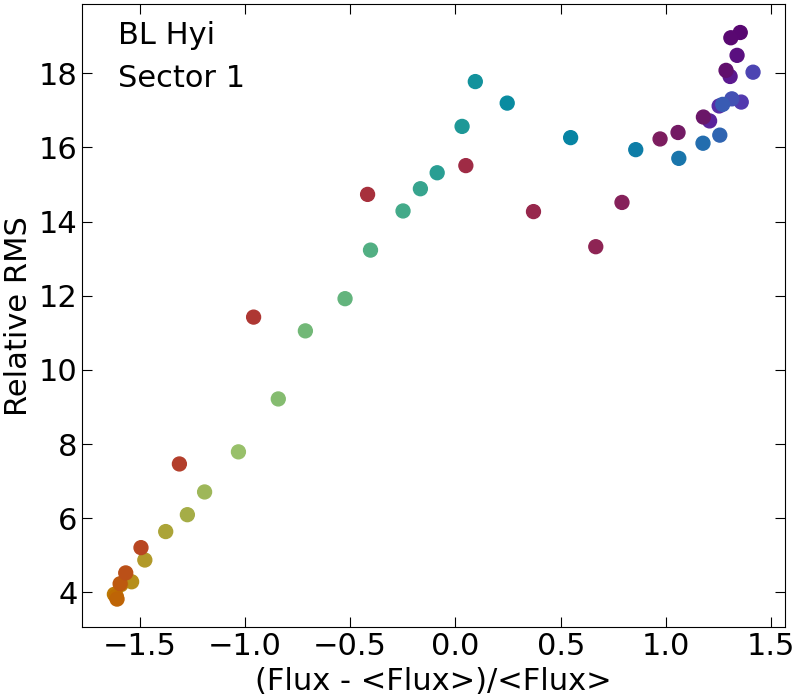}}
\resizebox{0.3\hsize}{!}{\includegraphics{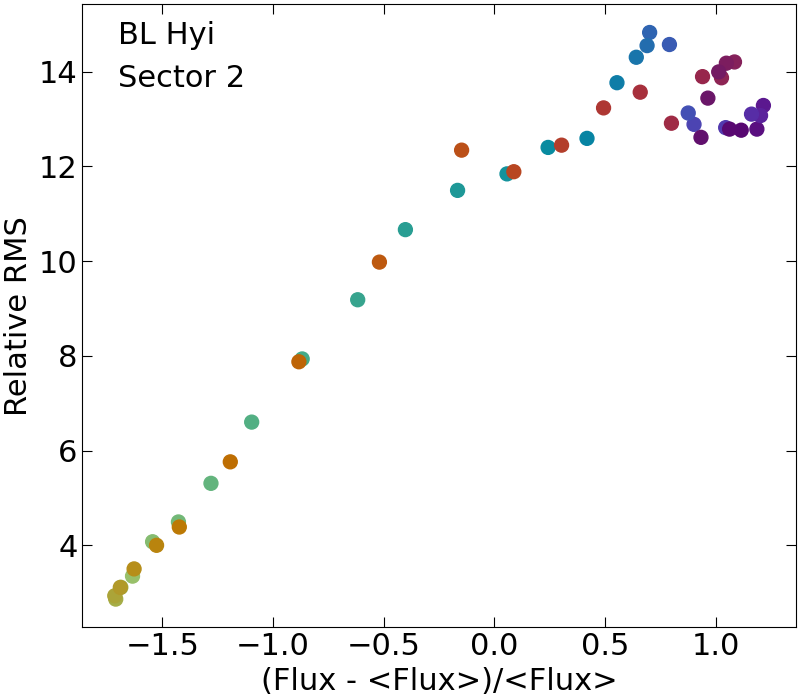}}
\resizebox{0.3\hsize}{!}{\includegraphics{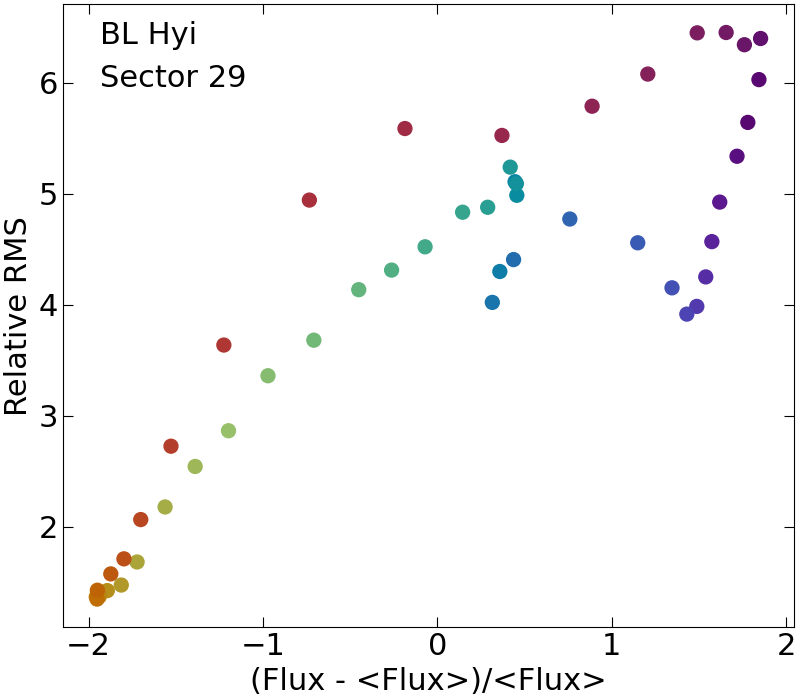}}\\
\resizebox{0.3\hsize}{!}{\includegraphics{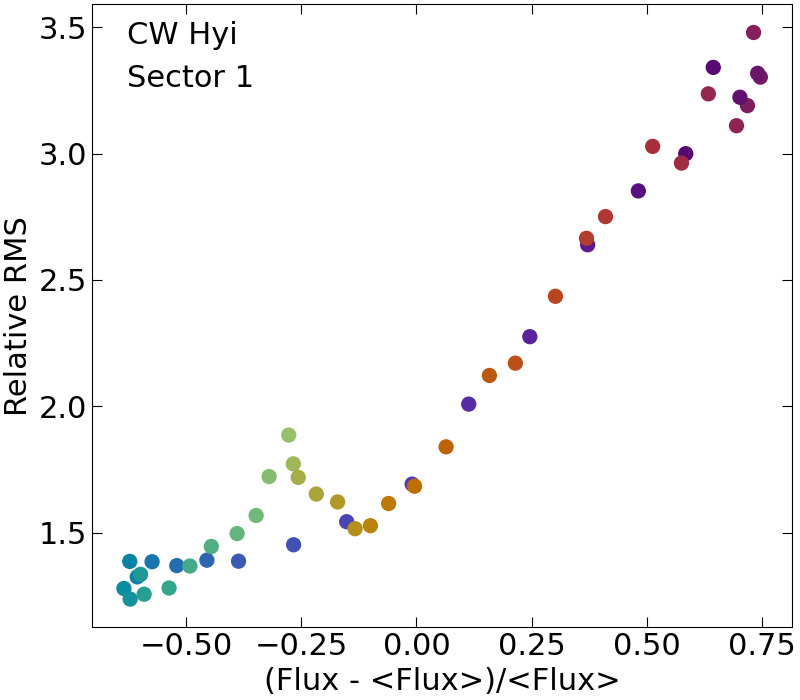}}
\resizebox{0.3\hsize}{!}{\includegraphics{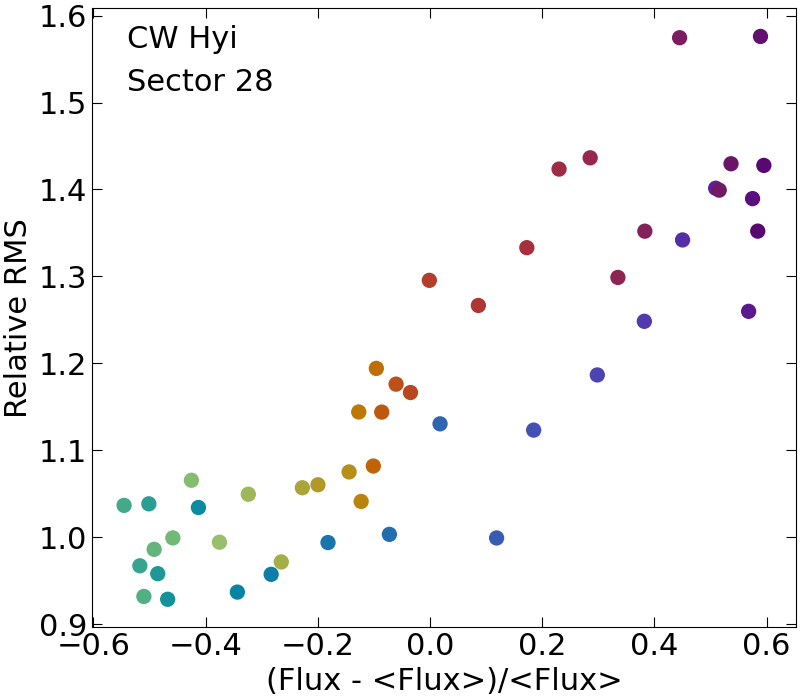}}
\resizebox{0.3\hsize}{!}{\includegraphics{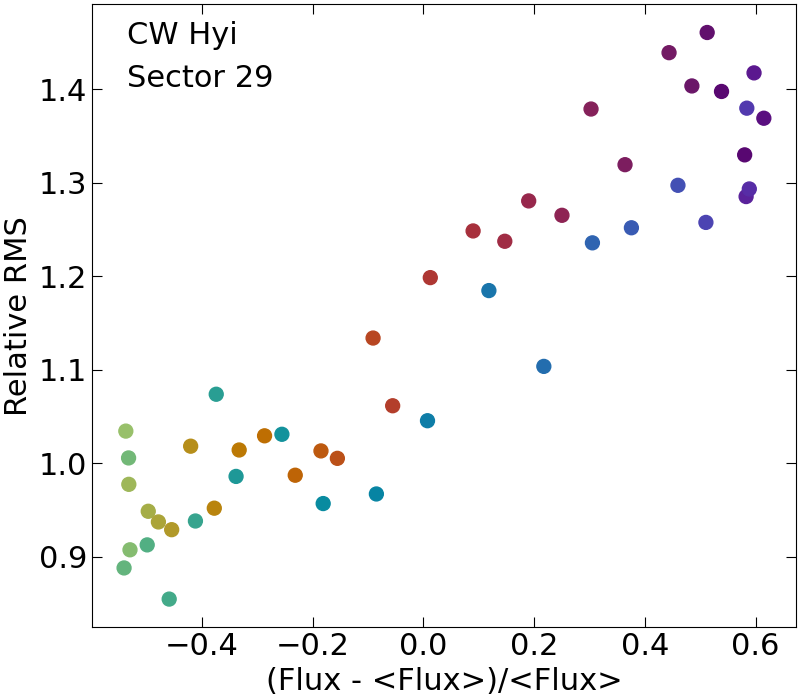}}\\
\resizebox{0.3\hsize}{!}{\includegraphics{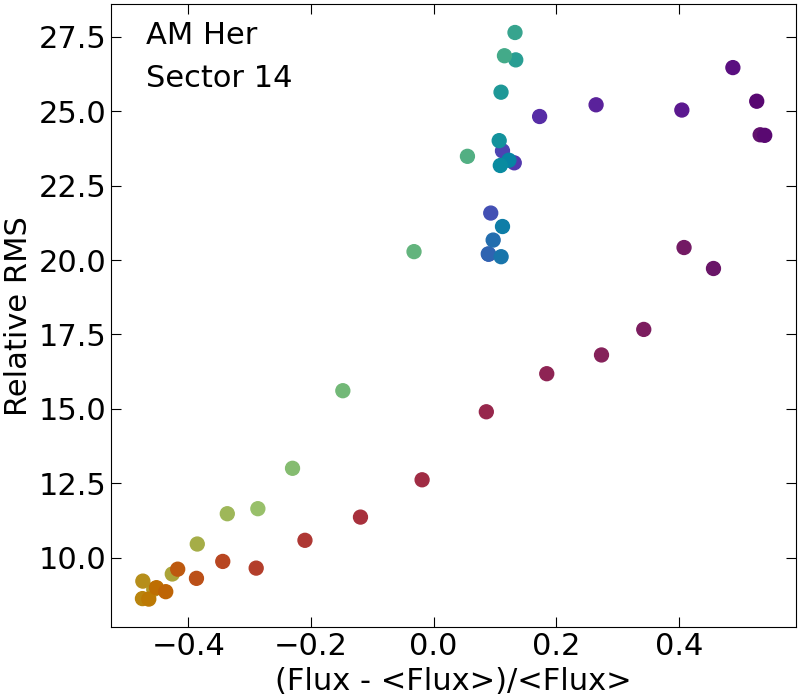}}
\resizebox{0.3\hsize}{!}{\includegraphics{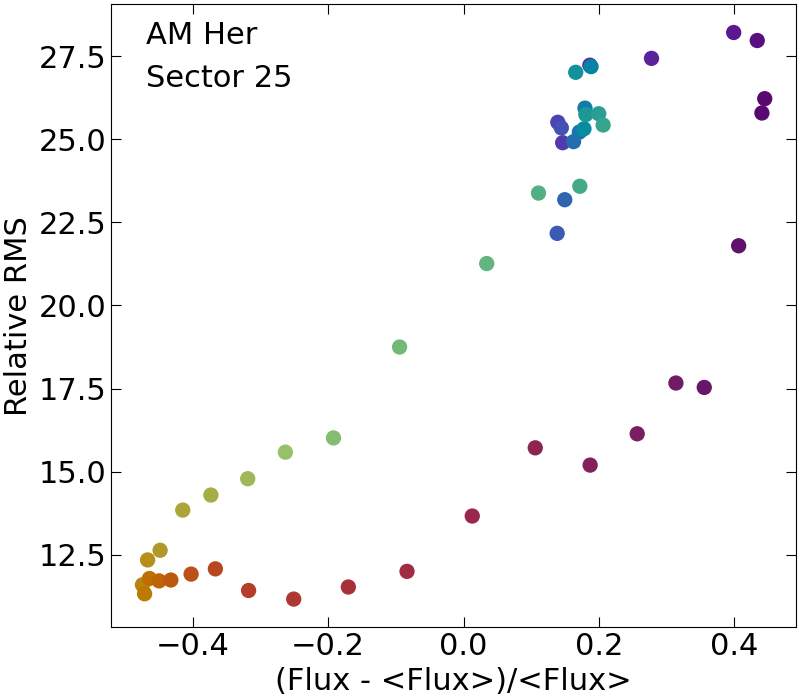}}
\resizebox{0.3\hsize}{!}{\includegraphics{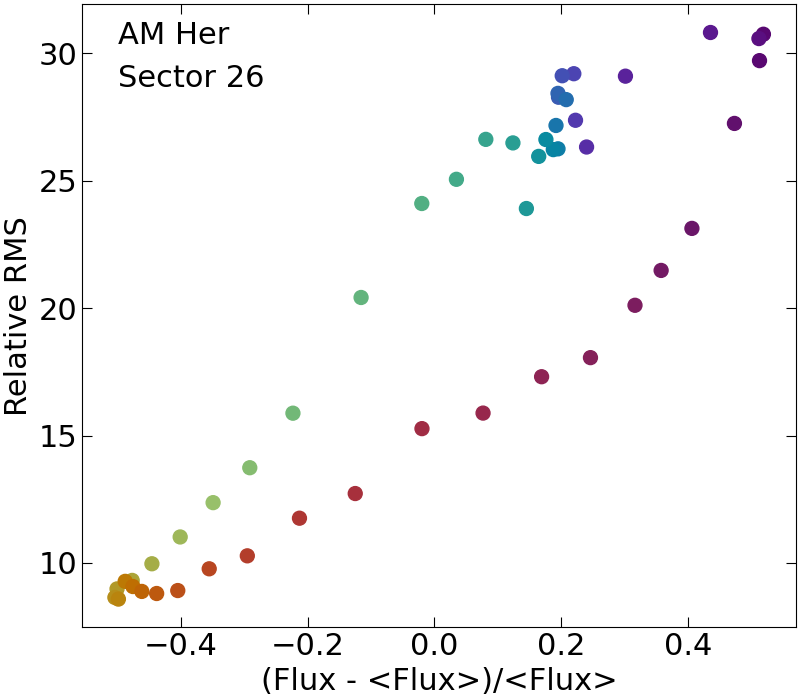}}\\
\resizebox{0.3\hsize}{!}{\includegraphics{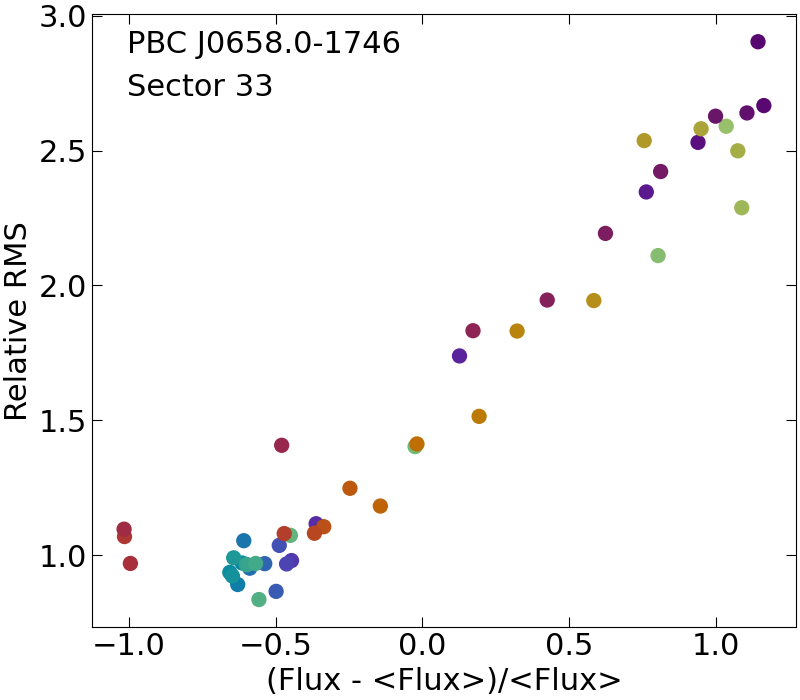}}
\resizebox{0.3\hsize}{!}{\includegraphics{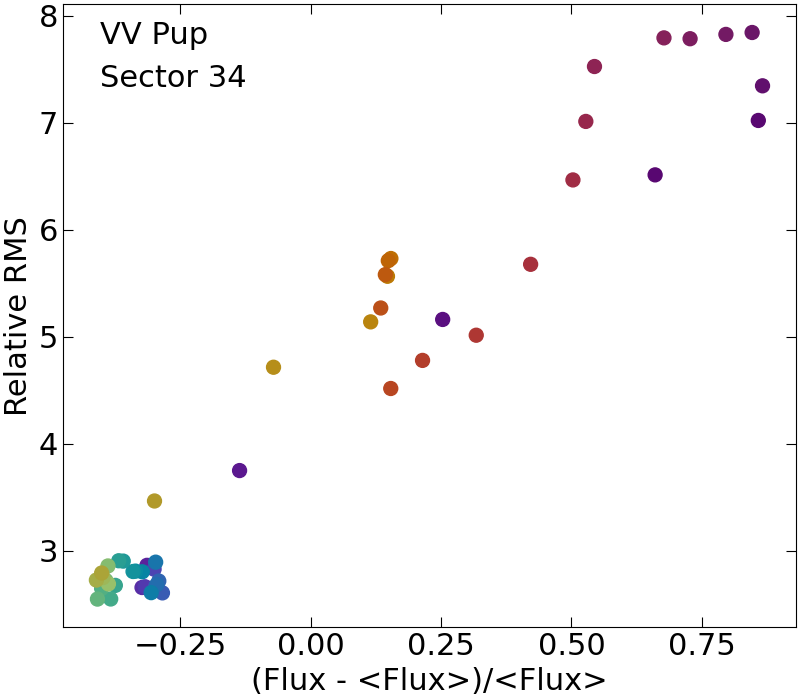}}
\resizebox{0.3\hsize}{!}{\includegraphics{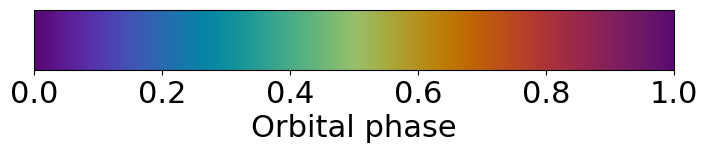}}


    \caption{The phased RMS-flux (PRF) relationships for polars, which demonstrates the relationship between the flickaring amplitude and the mean flux as a function of orbital phase. The orbital phase is marked with colour. For clarity we binned data in 50 bins equally distributed with the orbital period. }
    \label{fig:rms_flux_relation}
\end{figure*}

While the results of \citet{2021MNRAS.503..953B} could suggest that in polars amplitude of flickering during the orbital cycle should be correlated with changes in brightness, this was not always the case in our sample. In order to visualize this fact, we constructed orbital phase depend plots of RMS and mean flux (Fig.~\ref{fig:rms_flux_relation}), hereafter phased RMS-flux (PRF) relationships. MR~Ser, CTCV~J1928-5001, and PBC~J0658.0-1746 with exception of the time of eclipse clearly seem to follow a linear PRF relation. However, we stress that PRF relations are fundamentally different from RMS-flux relations commonly used to study flickering (e.g. \citealt{2005MNRAS.359..345U}). This is because PRF is binned in orbital phase and a linear PRF relationship simply visualizes a correlation between the mean light and amplitude of flickering. Therefore, PRF relations cannot be used to discuss the nature of flickering in a similar fashion as in \citet{2005MNRAS.359..345U}.

The most informative star in our sample is BL~Hyi, as during \textit{TESS} sector 29 two eclipses of the flickering source occurred. This implies that flickering in this star originates in at least two distinct locations. The PRF relation of BL~Hyi reflects this fact, which is best visible during \textit{TESS} sector 1. Namely, at orbital phases $\sim$0.4 and $\sim$0.9 the star switches between two linear PRF relations (Fig.~\ref{fig:rms_flux_relation}). These relations imply a higher flickering amplitude for a given flux between orbital phases $\sim$0.4 and $\sim$0.9. This is somewhat surprising, as it suggests that an additional flickering component is visible when the star brightness is lowest. A similar switch between two PRF relations occurs in CW~Hyi. Namely, during \textit{TESS} sector 1 between orbital phases $\sim$0.0 and $\sim$0.5, the relation is \ steeper.  Conversely to BL~Hyi, the switch between the two PRF relations in CW~Hyi is smooth. During  \textit{TESS} sectors 28 and 29, the flickering amplitude in CW~Hyi is too low to draw any conclusions.

The most puzzling PRF relation is observed in AM~Her, where a hysteresis-like dependent behavior is observed (Fig.~\ref{fig:rms_flux_relation}). Interestingly, at orbital phase $\sim$0.5 a switch between two PRF relations occurs without any changes in the mean brightness, which corresponds to the plateau in the light-curve (Fig.~\ref{fig:phased_rms_amher}). This switch is reflected by the fact that during \textit{TESS} sector 14 a plateau was accompanied by an increase in the flickering amplitude (Fig.~\ref{fig:phased_rms_amher}). A similar switch between two PRF relations during a plateau was observed in VV~Pup (Fig.~\ref{fig:phased_rms_vvpup},\ref{fig:rms_flux_relation}). The fact that both of these stars experienced a similar switch between two PRF relations during a plateau in the light-curve could be useful for future models of flickering in polars. A model of AM~Her  presented by \citet{2001A&A...372..557G} showed that during the plateau observed by \textit{TESS} a second peak in the light-curve due to cyclotron emission was expected. Since this peak was not present in the \textit{TESS} observations it is possible that it was eclipsed by the accretion stream. However, this peak was present in the variability of flickering amplitude, which would imply that flickering could originate as reprocessed cyclotron emission above the cyclotron emission place of origin. This is similar to suggestion of \citet{2021MNRAS.503..953B} that flickering is directly from cyclotron radiation near the post shock region.

\section{Discussion and conclusions}\label{sec:conclusion}

In this work, we analyzed the dependence of flickering amplitude on the orbital phase in a sample of seven polars. This is the biggest sample of polars studied with this method, where the previous samples included single objects \citep[e.g.][]{2021MNRAS.503..953B}.

We found that MR Ser, CTCV J1928-5001, and PBC J0658.0-1746 with exception of the time of eclipse followed the expected correlation between the brightness of the system and the flickering amplitude. However, the behavior of the remaining polars was different in each case. In some of the objects, the flickering source was eclipsed without any associated changes in brightness. In BL~Hyi two eclipses of the flickering source were observed during an orbital cycle, only one of which was accompanied by changes in the mean brightness. The results suggest the presence of at least two distinct flickering sources in some polars.

During the orbital cycle, some of the polars switch between two separate linear PRF relations.  Moreover, the source of flickering can be eclipsed during an orbital cycle even when there is no eclipse of the dominant light-source. Alternatively, in some of the objects, the dominant light-source could be eclipsed while the flickering source was not eclipsed. This could be the case in AM~Her, where the RMS-flux relation showed hysteresis-like behavior and a significant change in the flickering amplitude during the optical plateau. These results suggests that at least one flickering source originates in a large distance from the main light-source.   Moreover, the fact that some polars show one PRF relation and in some polars two separate linear PRF relations are present may suggest that each relation originates at different accretion pole. If confirmed, this may be used distinguish between one- and two-pole accreting white dwarfs.

While we showed that at least two flickering sources are present in some of our objects, it seems that only one can be located. Namely, we partially confirm the conclusion of \citet{2021MNRAS.503..953B} that flickering originates from a cyclotron or reprocessed X-ray radiation at the bottom of the accretion column. The second source of flickering remains elusive. However, the second source could be located when a detailed comparison to the models of polars similar to the one presented by \citet{2001A&A...372..557G} will be performed in the future. One of the possibilities that would ensure a large distance between the main light-source and location of flickering would be flickering generated by accretion of large diamagnetic blobs \citep{2002A&A...394..171H}.     If cyclotron radiation is a significant source of flickering, cyclotron beaming can influence the amplitude of flickering at a given orbital phase. This can be tested by studying the changes of flickering amplitude with orbital phase at different wavelengths. Since the effects of cyclotron beaming decreases at lower wavelengths its influence on changes in amplitude of flickering will be depended on the observed spectral range. However, a wavelength depended behaviour of flickering may be associated with different models and polarimetric observations may be needed to unravel the connection between flickering in polars and cyclotron radiation.

\section*{Acknowledgements}

This work was supported by STFC [ST/T000244/1] and Polish National Science Center grant 2021/40/C/ST9/00186.

\section*{DATA AVAILABILITY}
The data underlying this article is publicly available. The derived data generated in this work will be shared on a reasonable request to the corresponding author.





\bibliographystyle{mnras}
\bibliography{literature} 



\appendix


\bsp	
\label{lastpage}
\end{document}